%% file: CSforHS-ISIT.tex
\documentclass[conference,10pt,twocolumn,a4paper]{IEEEtran}

\usepackage[nosort]{cite}        
\usepackage[T1]{fontenc}
\usepackage{graphicx}

\usepackage{amssymb}
\usepackage[cmex10]{amsmath}
\usepackage{paralist}
\usepackage{functan}
\usepackage{url}

\usepackage{bbm}

\IEEEoverridecommandlockouts 

\input{vmr-symbols-vecbold}

\input{standard-macros}

\input{defs}

\renewcommand{\le}{\leqslant}
\renewcommand{\ge}{\geqslant}

\newcommand{\pmatspace}{\vspace{-0.1cm}}
\newcommand{\ppmatspace}{\vspace{0.1cm}}

\begin{document}

\title{Sparse Signal Recovery in Hilbert Spaces}
\author{
\IEEEauthorblockN{Graeme Pope and Helmut B\"olcskei}\\[-0.3cm]
\IEEEauthorblockA{Dept.~of IT \& EE, ETH Zurich, 8092 Zurich, Switzerland \\
Email: \{gpope,\ boelcskei\}@nari.ee.ethz.ch} 
 \thanks{The authors would like to thank C.~Aubel, R.~Heckel, R.~Pope, D.~Stotz, and C.~Studer  for inspiring discussions.}

}

\maketitle

\begin{abstract}
This paper reports an effort to consolidate numerous  coherence-based  sparse signal recovery results available in the literature.
We present a single theory that applies to general  Hilbert spaces with the sparsity of a signal  defined as the  number of (possibly infinite-dimensional) subspaces participating in the signal's representation.
Our general results recover uncertainty relations and  coherence-based  recovery thresholds   for sparse signals, block-sparse signals, multi-band signals, signals in shift-invariant spaces, and signals in  finite unions of (possibly infinite-dimensional) subspaces.
Moreover, we improve upon and generalize several of the existing results and, in many cases, we find shortened and simplified proofs.
\end{abstract}

\input intro.tex

\input sampling.tex
\input defns_ur.tex
\input recovery.tex
\input discussion.tex
\input uncertainty.tex

\bibliographystyle{IEEEtran} 

\bibliography{refs}

\end{document}

%% file: vmr-symbols-vecbold.tex
\usepackage{amssymb}
\usepackage{amsfonts}
\usepackage{mathrsfs}
\usepackage{xspace}
\usepackage{bm}
\usepackage{upgreek}

\newcommand{\safemath}[2]{\newcommand{#1}{\ensuremath{#2}\xspace}}

\renewcommand{\vec}[1]{\boldsymbol{#1}}
\newcommand{\mat}[1]{\boldsymbol{#1}}

\safemath{\bma}{\boldsymbol{a}}
\safemath{\bmb}{\boldsymbol{b}}
\safemath{\bmc}{\boldsymbol{c}}
\safemath{\bmd}{\boldsymbol{d}}
\safemath{\bme}{\boldsymbol{e}}
\safemath{\bmf}{\boldsymbol{f}}
\safemath{\bmg}{\boldsymbol{g}}
\safemath{\bmh}{\boldsymbol{h}}
\safemath{\bmi}{\boldsymbol{i}}
\safemath{\bmj}{\boldsymbol{j}}
\safemath{\bmk}{\boldsymbol{k}}
\safemath{\bml}{\boldsymbol{l}}
\safemath{\bmm}{\boldsymbol{m}}
\safemath{\bmn}{\boldsymbol{n}}
\safemath{\bmo}{\boldsymbol{o}}
\safemath{\bmp}{\boldsymbol{p}}
\safemath{\bmq}{\boldsymbol{q}}
\safemath{\bmr}{\boldsymbol{r}}
\safemath{\bms}{\boldsymbol{s}}
\safemath{\bmt}{\boldsymbol{t}}
\safemath{\bmu}{\boldsymbol{u}}
\newcommand{\bmv}{\boldsymbol{v}}
\safemath{\bmw}{\boldsymbol{w}}
\safemath{\bmx}{\boldsymbol{x}}
\safemath{\bmy}{\boldsymbol{y}}
\safemath{\bmz}{\boldsymbol{z}}
\safemath{\bmzero}{\boldsymbol{0}}
\safemath{\bmone}{\boldsymbol{1}}

\bmdefine{\biad}{a}
\bmdefine{\bibd}{b}
\bmdefine{\bicd}{c}
\bmdefine{\bidd}{d}
\bmdefine{\bied}{e}
\bmdefine{\bifd}{f}
\bmdefine{\bigd}{g}
\bmdefine{\bihd}{h}
\bmdefine{\biid}{i}
\bmdefine{\bijd}{j}
\bmdefine{\bikd}{k}
\bmdefine{\bild}{l}
\bmdefine{\bimd}{m}
\bmdefine{\bind}{n}
\bmdefine{\biod}{o}
\bmdefine{\bipd}{p}
\bmdefine{\biqd}{q}
\bmdefine{\bird}{r}
\bmdefine{\bisd}{s}
\bmdefine{\bitd}{t}
\bmdefine{\biud}{u}
\bmdefine{\bivd}{v}
\bmdefine{\biwd}{w}
\bmdefine{\bixd}{x}
\bmdefine{\biyd}{y}
\bmdefine{\bizd}{z}

\bmdefine{\bixid}{\xi}
\bmdefine{\bilambdad}{\lambda}
\bmdefine{\bimud}{\mu}
\bmdefine{\bithetad}{\theta}
\bmdefine{\biphid}{\phi}
\bmdefine{\bideltad}{\delta}

\safemath{\bmia}{\biad}
\safemath{\bmib}{\bibd}
\safemath{\bmic}{\bicd}
\safemath{\bmid}{\bidd}
\safemath{\bmie}{\bied}
\safemath{\bmif}{\bifd}
\safemath{\bmig}{\bigd}
\safemath{\bmih}{\bihd}
\safemath{\bmii}{\biid}
\safemath{\bmij}{\bijd}
\safemath{\bmik}{\bikd}
\safemath{\bmil}{\bild}
\safemath{\bmim}{\bimd}
\safemath{\bmin}{\bind}
\safemath{\bmio}{\biod}
\safemath{\bmip}{\bipd}
\safemath{\bmiq}{\biqd}
\safemath{\bmir}{\bird}
\safemath{\bmis}{\bisd}
\safemath{\bmit}{\bitd}
\safemath{\bmiu}{\biud}
\safemath{\bmiv}{\bivd}
\safemath{\bmiw}{\biwd}
\safemath{\bmix}{\bixd}
\safemath{\bmiy}{\biyd}
\safemath{\bmiz}{\bizd}

\safemath{\bmxi}{\bixid}
\safemath{\bmlambda}{\bilambdad}
\safemath{\bmmu}{\bimud}
\safemath{\bmtheta}{\bithetad}
\safemath{\bmphi}{\biphid}
\safemath{\bmdelta}{\bideltad}

\safemath{\bA}{\mathbf{A}}
\safemath{\bB}{\mathbf{B}}
\safemath{\bC}{\mathbf{C}}
\safemath{\bD}{\mathbf{D}}
\safemath{\bE}{\mathbf{E}}
\safemath{\bF}{\mathbf{F}}
\safemath{\bG}{\mathbf{G}}
\safemath{\bH}{\mathbf{H}}
\safemath{\bI}{\mathbf{I}}
\safemath{\bJ}{\mathbf{J}}
\safemath{\bK}{\mathbf{K}}
\safemath{\bL}{\mathbf{L}}
\safemath{\bM}{\mathbf{M}}
\safemath{\bN}{\mathbf{N}}
\safemath{\bO}{\mathbf{O}}
\safemath{\bP}{\mathbf{P}}
\safemath{\bQ}{\mathbf{Q}}
\safemath{\bR}{\mathbf{R}}
\safemath{\bS}{\mathbf{S}}
\safemath{\bT}{\mathbf{T}}
\safemath{\bU}{\mathbf{U}}
\safemath{\bV}{\mathbf{V}}
\safemath{\bW}{\mathbf{W}}
\safemath{\bX}{\mathbf{X}}
\safemath{\bY}{\mathbf{Y}}
\safemath{\bZ}{\mathbf{Z}}

\safemath{\bzz}{\boldsymbol{0}}
\safemath{\bOne}{\boldsymbol{1}}
\safemath{\bDelta}{\boldsymbol{\Delta}}
\safemath{\bLambda}{\boldsymbol{\UpLambda}}
\safemath{\bPhi}{\boldsymbol{\Upphi}}
\safemath{\bSigma}{\boldsymbol{\Upsigma}}
\safemath{\bOmega}{\boldsymbol{\Upomega}}
\safemath{\bTheta}{\boldsymbol{\Uptheta}}

\bmdefine{\biAd}{A}
\bmdefine{\biBd}{B}
\bmdefine{\biCd}{C}
\bmdefine{\biDd}{D}
\bmdefine{\biEd}{E}
\bmdefine{\biFd}{F}
\bmdefine{\biGd}{G}
\bmdefine{\biHd}{H}
\bmdefine{\biId}{I}
\bmdefine{\biJd}{J}
\bmdefine{\biKd}{K}
\bmdefine{\biLd}{L}
\bmdefine{\biMd}{M}
\bmdefine{\biOd}{N}
\bmdefine{\biPd}{O}
\bmdefine{\biQd}{P}
\bmdefine{\biRd}{R}
\bmdefine{\biSd}{S}
\bmdefine{\biTd}{T}
\bmdefine{\biUd}{U}
\bmdefine{\biVd}{V}
\bmdefine{\biWd}{W}
\bmdefine{\biXd}{X}
\bmdefine{\biYd}{Y}
\bmdefine{\biZd}{Z}

\bmdefine{\biDelta}{\Delta}
\bmdefine{\biLambda}{\Lambda}
\bmdefine{\biPhi}{\Phi}
\bmdefine{\biSigma}{\Sigma}
\bmdefine{\biOmega}{\Omega}
\bmdefine{\biTheta}{\Theta}

\safemath{\bimA}{\biAd}
\safemath{\bimB}{\biBd}
\safemath{\bimC}{\biCd}
\safemath{\bimD}{\biDd}
\safemath{\bimE}{\biEd}
\safemath{\bimF}{\biFd}
\safemath{\bimG}{\biGd}
\safemath{\bimH}{\biHd}
\safemath{\bimI}{\biId}
\safemath{\bimJ}{\biJd}
\safemath{\bimK}{\biKd}
\safemath{\bimL}{\biLd}
\safemath{\bimM}{\biMd}
\safemath{\bimN}{\biNd}
\safemath{\bimO}{\biOd}
\safemath{\bimP}{\biPd}
\safemath{\bimQ}{\biQd}
\safemath{\bimR}{\biRd}
\safemath{\bimS}{\biSd}
\safemath{\bimT}{\biTd}
\safemath{\bimU}{\biUd}
\safemath{\bimV}{\biVd}
\safemath{\bimW}{\biWd}
\safemath{\bimX}{\biXd}
\safemath{\bimY}{\biYd}
\safemath{\bimZ}{\biZd}

\safemath{\bimDelta}{\biDelta}
\safemath{\bimLambda}{\biLambda}
\safemath{\bimPhi}{\biPhi}
\safemath{\bimSigma}{\biSigma}
\safemath{\bimOmega}{\biOmega}
\safemath{\bimTheta}{\biTheta}

\safemath{\setA}{\mathcal{A}}
\safemath{\setB}{\mathcal{B}}
\safemath{\setC}{\mathcal{C}}
\safemath{\setD}{\mathcal{D}}
\safemath{\setE}{\mathcal{E}}
\safemath{\setF}{\mathcal{F}}
\safemath{\setG}{\mathcal{G}}
\safemath{\setH}{\mathcal{H}}
\safemath{\setI}{\mathcal{I}}
\safemath{\setJ}{\mathcal{J}}
\safemath{\setK}{\mathcal{K}}
\safemath{\setL}{\mathcal{L}}
\safemath{\setM}{\mathcal{M}}
\safemath{\setN}{\mathcal{N}}
\safemath{\setO}{\mathcal{O}}
\safemath{\setP}{\mathcal{P}}
\safemath{\setQ}{\mathcal{Q}}
\safemath{\setR}{\mathcal{R}}
\safemath{\setS}{\mathcal{S}}
\safemath{\setT}{\mathcal{T}}
\safemath{\setU}{\mathcal{U}}
\safemath{\setV}{\mathcal{V}}
\safemath{\setW}{\mathcal{W}}
\safemath{\setX}{\mathcal{X}}
\safemath{\setY}{\mathcal{Y}}
\safemath{\setZ}{\mathcal{Z}}
\safemath{\emptySet}{\varnothing}

\safemath{\colA}{\mathscr{A}}
\safemath{\colB}{\mathscr{B}}
\safemath{\colC}{\mathscr{C}}
\safemath{\colD}{\mathscr{D}}
\safemath{\colE}{\mathscr{E}}
\safemath{\colF}{\mathscr{F}}
\safemath{\colG}{\mathscr{G}}
\safemath{\colH}{\mathscr{H}}
\safemath{\colI}{\mathscr{I}}
\safemath{\colJ}{\mathscr{J}}
\safemath{\colK}{\mathscr{K}}
\safemath{\colL}{\mathscr{L}}
\safemath{\colM}{\mathscr{M}}
\safemath{\colN}{\mathscr{N}}
\safemath{\colO}{\mathscr{O}}
\safemath{\colP}{\mathscr{P}}
\safemath{\colQ}{\mathscr{Q}}
\safemath{\colR}{\mathscr{R}}
\safemath{\colS}{\mathscr{S}}
\safemath{\colT}{\mathscr{T}}
\safemath{\colU}{\mathscr{U}}
\safemath{\colV}{\mathscr{V}}
\safemath{\colW}{\mathscr{W}}
\safemath{\colX}{\mathscr{X}}
\safemath{\colY}{\mathscr{Y}}
\safemath{\colZ}{\mathscr{Z}}

\safemath{\opA}{\mathbb{A}}
\safemath{\opB}{\mathbb{B}}
\safemath{\opC}{\mathbb{C}}
\safemath{\opD}{\mathbb{D}}
\safemath{\opE}{\mathbb{E}}
\safemath{\opF}{\mathbb{F}}
\safemath{\opG}{\mathbb{G}}
\safemath{\opH}{\mathbb{H}}
\safemath{\opI}{\mathbb{I}}
\safemath{\opJ}{\mathbb{J}}
\safemath{\opK}{\mathbb{K}}
\safemath{\opL}{\mathbb{L}}
\safemath{\opM}{\mathbb{M}}
\safemath{\opN}{\mathbb{N}}
\safemath{\opO}{\mathbb{O}}
\safemath{\opP}{\mathbb{P}}
\safemath{\opQ}{\mathbb{Q}}
\safemath{\opR}{\mathbb{R}}
\safemath{\opS}{\mathbb{S}}
\safemath{\opT}{\mathbb{T}}
\safemath{\opU}{\mathbb{U}}
\safemath{\opV}{\mathbb{V}}
\safemath{\opW}{\mathbb{W}}
\safemath{\opX}{\mathbb{X}}
\safemath{\opY}{\mathbb{Y}}
\safemath{\opZ}{\mathbb{Z}}
\safemath{\opZero}{\mathbb{O}}
\safemath{\identityop}{\opI}

\safemath{\veca}{\bma}
\safemath{\vecb}{\bmb}
\safemath{\vecc}{\bmc}
\safemath{\vecd}{\bmd}
\safemath{\vece}{\bme}
\safemath{\vecf}{\bmf}
\safemath{\vecg}{\bmg}
\safemath{\vech}{\bmh}
\safemath{\veci}{\bmi}
\safemath{\vecj}{\bmj}
\safemath{\veck}{\bmk}
\safemath{\vecl}{\bml}
\safemath{\vecm}{\bmm}
\safemath{\vecn}{\bmn}
\safemath{\veco}{\bmo}
\safemath{\vecp}{\bmp}
\safemath{\vecq}{\bmq}
\safemath{\vecr}{\bmr}
\safemath{\vecs}{\bms}
\safemath{\vect}{\bmt}
\safemath{\vecu}{\bmu}
\safemath{\vecv}{\bmv}
\safemath{\vecw}{\bmw}
\safemath{\vecx}{\bmx}
\safemath{\vecy}{\bmy}
\safemath{\vecz}{\bmz}

\safemath{\va}{\bma}
\safemath{\vb}{\bmb}
\safemath{\vc}{\bmc}
\safemath{\vd}{\bmd}
\safemath{\ve}{\bme}
\safemath{\vf}{\bmf}
\safemath{\vg}{\bmg}
\safemath{\vh}{\bmh}
\safemath{\vi}{\bmi}
\safemath{\vj}{\bmj}
\safemath{\vk}{\bmk}
\safemath{\vl}{\bml}
\safemath{\vm}{\bmm}
\safemath{\vn}{\bmn}
\safemath{\vo}{\bmo}
\safemath{\vp}{\bmp}
\safemath{\vq}{\bmq}
\safemath{\vr}{\bmr}
\safemath{\vs}{\bms}
\safemath{\vt}{\bmt}
\safemath{\vu}{\bmu}
\safemath{\vv}{\bmv}
\safemath{\vw}{\bmw}
\safemath{\vx}{\bmx}
\safemath{\vy}{\bmy}
\safemath{\vz}{\bmz}

\safemath{\veczero}{\bmzero}
\safemath{\vecone}{\bmone}
\safemath{\vecxi}{\bmxi}
\safemath{\veclambda}{\bmlambda}
\safemath{\vecmu}{\bmmu}
\safemath{\vectheta}{\bmtheta}
\safemath{\vecphi}{\bmphi}
\safemath{\vecdelta}{\bmdelta}

\safemath{\matA}{\bA}
\safemath{\matB}{\bB}
\safemath{\matC}{\bC}
\safemath{\matD}{\bD}
\safemath{\matE}{\bE}
\safemath{\matF}{\bF}
\safemath{\matG}{\bG}
\safemath{\matH}{\bH}
\safemath{\matI}{\bI}
\safemath{\matJ}{\bJ}
\safemath{\matK}{\bK}
\safemath{\matL}{\bL}
\safemath{\matM}{\bM}
\safemath{\matN}{\bN}
\safemath{\matO}{\bO}
\safemath{\matP}{\bP}
\safemath{\matQ}{\bQ}
\safemath{\matR}{\bR}
\safemath{\matS}{\bS}
\safemath{\matT}{\bT}
\safemath{\matU}{\bU}
\safemath{\matV}{\bV}
\safemath{\matW}{\bW}
\safemath{\matX}{\bX}
\safemath{\matY}{\bY}
\safemath{\matZ}{\bZ}
\safemath{\matzero}{\bmzero}

\safemath{\matDelta}{\bDelta}
\safemath{\matLambda}{\bLambda}
\safemath{\matPhi}{\bPhi}
\safemath{\matSigma}{\bSigma}
\safemath{\matOmega}{\bOmega}
\safemath{\matTheta}{\bTheta}

\safemath{\matidentity}{\matI}
\safemath{\matone}{\matO}

\safemath{\rnda}{A}
\safemath{\rndb}{B}
\safemath{\rndc}{C}
\safemath{\rndd}{D}
\safemath{\rnde}{E}
\safemath{\rndf}{F}
\safemath{\rndg}{G}
\safemath{\rndh}{H}
\safemath{\rndi}{I}
\safemath{\rndj}{J}
\safemath{\rndk}{K}
\safemath{\rndl}{L}
\safemath{\rndm}{M}
\safemath{\rndn}{N}
\safemath{\rndo}{O}
\safemath{\rndp}{P}
\safemath{\rndq}{Q}
\safemath{\rndr}{R}
\safemath{\rnds}{S}
\safemath{\rndt}{T}
\safemath{\rndu}{U}
\safemath{\rndv}{V}
\safemath{\rndw}{W}
\safemath{\rndx}{X}
\safemath{\rndy}{Y}
\safemath{\rndz}{Z}

\safemath{\rveca}{\bimA}
\safemath{\rvecb}{\bimB}
\safemath{\rvecc}{\bimC}
\safemath{\rvecd}{\bimD}
\safemath{\rvece}{\bimE}
\safemath{\rvecf}{\bimF}
\safemath{\rvecg}{\bimG}
\safemath{\rvech}{\bimH}
\safemath{\rveci}{\bimI}
\safemath{\rvecj}{\bimJ}
\safemath{\rveck}{\bimK}
\safemath{\rvecl}{\bimL}
\safemath{\rvecm}{\bimM}
\safemath{\rvecn}{\bimN}
\safemath{\rveco}{\bomO}
\safemath{\rvecp}{\bimP}
\safemath{\rvecq}{\bimQ}
\safemath{\rvecr}{\bimR}
\safemath{\rvecs}{\bimS}
\safemath{\rvect}{\bimT}
\safemath{\rvecu}{\bimU}
\safemath{\rvecv}{\bimV}
\safemath{\rvecw}{\bimW}
\safemath{\rvecx}{\bimX}
\safemath{\rvecy}{\bimY}
\safemath{\rvecz}{\bimZ}

\safemath{\rvecxi}{\bmxi}
\safemath{\rveclambda}{\bmlambda}
\safemath{\rvecmu}{\bmmu}
\safemath{\rvectheta}{\bmtheta}
\safemath{\rvecphi}{\bmphi}

\safemath{\rmatA}{\bimA}
\safemath{\rmatB}{\bimB}
\safemath{\rmatC}{\bimC}
\safemath{\rmatD}{\bimD}
\safemath{\rmatE}{\bimE}
\safemath{\rmatF}{\bimF}
\safemath{\rmatG}{\bimG}
\safemath{\rmatH}{\bimH}
\safemath{\rmatI}{\bimI}
\safemath{\rmatJ}{\bimJ}
\safemath{\rmatK}{\bimK}
\safemath{\rmatL}{\bimL}
\safemath{\rmatM}{\bimM}
\safemath{\rmatN}{\bimN}
\safemath{\rmatO}{\bimO}
\safemath{\rmatP}{\bimP}
\safemath{\rmatQ}{\bimQ}
\safemath{\rmatR}{\bimR}
\safemath{\rmatS}{\bimS}
\safemath{\rmatT}{\bimT}
\safemath{\rmatU}{\bimU}
\safemath{\rmatV}{\bimV}
\safemath{\rmatW}{\bimW}
\safemath{\rmatX}{\bimX}
\safemath{\rmatY}{\bimY}
\safemath{\rmatZ}{\bimZ}

\safemath{\rmatDelta}{\bimDelta}
\safemath{\rmatLambda}{\bimLambda}
\safemath{\rmatPhi}{\bimPhi}
\safemath{\rmatSigma}{\bimSigma}
\safemath{\rmatOmega}{\bimOmega}
\safemath{\rmatTheta}{\bimTheta}

%% file: standard-macros.tex
\usepackage{amssymb}
\usepackage{amsfonts}
\usepackage{mathrsfs}
\usepackage{xspace}
\usepackage{bm}
\usepackage{fancyref}
\usepackage{textcomp}

\usepackage{multirow}
\usepackage{stmaryrd}

\usepackage{functan}

\newcommand{\lefto}{\mathopen{}\left}

\newcommand{\sub}[1]{\ensuremath{_{\subtext{#1}}}} 

\DeclareMathOperator*{\argmin}{arg\;min}		
\DeclareMathOperator*{\argmax}{arg\;max}		
\DeclareMathOperator*{\esssup}{ess\;sup}			
\DeclareMathOperator{\spark}{spark}

\DeclareMathOperator*{\mini}{minimize\ }
\DeclareMathOperator{\st}{subject\ to\ }

\newcommand{\abs}[1]{\lefto\lvert#1\right\rvert}		

\safemath{\dirac}{\delta}					
\safemath{\krond}{\dirac}					

\newcommand{\set}[1]{\left\{#1\right\}}		
\safemath{\upto}{\uparrow}
\safemath{\downto}{\downarrow}
\safemath{\iu}{j}							
\safemath{\hilseqspace}{l^{2}}				
\newcommand{\banachfunspace}[1]{\setL^{#1}}	
\safemath{\hilfunspace}{\banachfunspace{2}}	

\Macro{L2}{\mathrm{L}^2(\Omega)}
\Macro{li}{\infty}
\Macro{l0}{0}
\Macro{l1}{1}
\Macro{l2}{2}
\Macro{l20}{2,0}
\Macro{lp}{p}
\Macro{lq}{q}
\Macro{lpd}{p'}
\Macro{lqd}{q'}
\Macro{max}{\max}
\Macro{l20}{2,0}
\Macro{l21}{2,1}
\Macro{lpq}{p,q}
\Macro{lpqd}{p',q'}
\Macro{lpp}{p,p}
\Macro{lp0}{p,0}
\Macro{l10}{1,0}
\Macro{l2i}{2,\infty}
\Macro{F}{\text{F}}
\Macro{H}{\mathscr{H}}
\Macro{Ha}{\Ha}
\Macro{Hb}{\Hb}
\Macro{HaHb}{\Ha \rightarrow \Hb}
\Macro{HaHa}{\Ha \rightarrow \Ha}
\Macro{HHa}{\Ha,\Ha}

\Macro{Ha1}{\Ha,1}
\Macro{Ha0}{\Ha,0}
\Macro{H1}{1}

\Macro{l2l1}{2 \rightarrow 1}
\Macro{l2l2}{2 \rightarrow 2}

\safemath{\normal}{\mathcal{N}}			
\safemath{\jpg}{\mathcal{CN}}			
\safemath{\mchain}{\leftrightarrow}		

\safemath{\dB}{\,\mathrm{dB}}
\safemath{\dBm}{\,\mathrm{dBm}}
\safemath{\Hz}{\,\mathrm{Hz}}
\safemath{\kHz}{\,\mathrm{kHz}}
\safemath{\MHz}{\,\mathrm{MHz}}
\safemath{\GHz}{\,\mathrm{GHz}}
\safemath{\s}{\,\mathrm{s}}
\safemath{\ms}{\,\mathrm{ms}}
\safemath{\mus}{\,\mathrm{\text{\textmu}s}}
\safemath{\ns}{\,\mathrm{ns}}
\safemath{\ps}{\,\mathrm{ps}}
\safemath{\meter}{\,\mathrm{m}}
\safemath{\mm}{\,\mathrm{mm}}
\safemath{\cm}{\,\mathrm{cm}}
\safemath{\W}{\,\mathrm{W}}
\safemath{\mW}{\, \mathrm{mW}}
\safemath{\J}{\,\mathrm{J}}
\safemath{\K}{\,\mathrm{K}}
\safemath{\bit}{\,\mathrm{bit}}
\safemath{\nat}{\,\mathrm{nat}}

\safemath{\define}{\triangleq}			

\safemath{\equivalent}{\sim}
\safemath{\distas}{\sim}					
\safemath{\sdiff}{\Delta}				

\safemath{\reals}{\mathbb{R}}
\safemath{\positivereals}{\reals_{+}}
\safemath{\integers}{\mathbb{Z}}
\safemath{\posint}{\integers_{+}}
\safemath{\naturals}{\mathbb{N}}
\safemath{\posnaturals}{\naturals_{+}}
\safemath{\complexset}{\mathbb{C}}
\safemath{\rationals}{\mathbb{Q}}
\safemath{\lt}{\ell_2}
\safemath{\Lt}{\mathrm{L}_2}

\newcommand*{\fancyrefapplabelprefix}{app}		
\newcommand*{\fancyrefthmlabelprefix}{thm}		
\newcommand*{\fancyreflemlabelprefix}{lem}		
\newcommand*{\fancyrefcorlabelprefix}{cor}		
\newcommand*{\fancyrefdeflabelprefix}{def}		
\newcommand*{\fancyrefproplabelprefix}{prop}		
\newcommand*{\fancyrefexmpllabelprefix}{exmpl}
\frefformat{vario}{\fancyrefseclabelprefix}{Sec.~#1}
\frefformat{vario}{\fancyrefthmlabelprefix}{Theorem~#1}
\frefformat{vario}{\fancyreflemlabelprefix}{Lem.~#1}
\frefformat{vario}{\fancyrefcorlabelprefix}{Cor.~#1}
\frefformat{vario}{\fancyrefdeflabelprefix}{Definition~#1}
\frefformat{vario}{\fancyreffiglabelprefix}{Fig.~#1}
\frefformat{vario}{\fancyrefapplabelprefix}{App.~#1} 
\frefformat{vario}{\fancyrefeqlabelprefix}{(#1)}
\frefformat{vario}{\fancyrefproplabelprefix}{Prop.~#1}
\frefformat{vario}{\fancyrefexmpllabelprefix}{Example~#1}

%% file: defs.tex
 \newtheorem{thm}{Theorem}
 
 \newtheorem{defn}{Definition}

\safemath{\cplxi}{\imath}
\safemath{\cplxj}{\jmath}

\newcommand{\adj}{{H}}

\renewcommand{\it}[2]{{#1}_{#2}}

\newcommand{\pos}[1]{\lefto[#1\right]^{+}}

\renewcommand{\b}[2]{#1 \lefto[ #2 \right]}

\safemath{\Ha}{{\mathscr{H}}}
\safemath{\Hb}{{\mathscr{G}}}

\safemath{\proj}{\pi}
\safemath{\inc}{\imath}

\safemath{\eps}{\varepsilon}
\safemath{\op}{\varphi}

\safemath{\HBP}{\mathscr{H}\text{-}\textsc{BP}\xspace}
\safemath{\HPZ}{\mathscr{H}\text{-}\textsc{P0}\xspace}
\safemath{\Homp}{\mathscr{H}\text{-}\textsc{OMP}\xspace}
\safemath{\PZ}{\textsc{P0}\xspace}
\safemath{\BP}{\textsc{BP}\xspace}
\safemath{\OMP}{\textsc{OMP}\xspace}
\safemath{\BOMP}{\textsc{BOMP}\xspace}
\safemath{\BBP}{\textsc{Block-MP}\xspace}
\newcommand{\Gersgorin}{Ger\v{s}gorin\xspace}

\safemath{\lspread}{\kappa}

\renewcommand{\sub}[2]{#1^{(#2)}}

\safemath{\samp}{\Phi}
\safemath{\sampb}{\Psi}
\safemath{\sampcol}{\varphi}
\safemath{\subop}{\mathbb{T}}
\safemath{\sampbcol}{\psi}
\safemath{\dict}{\matD}
\safemath{\dcol}{\vec{d}}
\safemath{\ecol}{\vec{e}}

\safemath{\acol}{\vec{a}}
\safemath{\bcol}{\vec{b}}
\safemath{\partition}{\mathscr{P}}

\safemath{\sigv}{\vec{v}}
\safemath{\obsv}{\vec{z}}

\safemath{\sig}{v}
\safemath{\siga}{v}
\safemath{\sigsup}{\setV}
\safemath{\nsig}{n_v}
\safemath{\sigb}{u}
\safemath{\sigbsup}{\setU}
\safemath{\nsigb}{n_u}
\safemath{\out}{z}
\safemath{\obs}{z}
\safemath{\err}{e}
\safemath{\nerr}{n_e}
\safemath{\errsup}{\setE}

\safemath{\cb}{\delta}

\safemath{\dima}{N}
\safemath{\dimb}{M}
\safemath{\dimn}{\dima}
\safemath{\dimm}{\dimb}
\safemath{\dims}{S}
\safemath{\subs}{s}
\safemath{\subn}{n}
\safemath{\subna}{{n_1}}
\safemath{\subnb}{{n_2}}
\safemath{\subm}{m}
\safemath{\subdim}{d}

\safemath{\cdim}{\complexset^{\dimb \times \dima}}

\safemath{\infdim}{\setN_\infty} 
\safemath{\findim}{\setN_f}      

\safemath{\sia}{\phi}
\safemath{\sib}{\theta}
\safemath{\Na}{N_{\sia}}
\safemath{\Nb}{N_{\sib}}
\safemath{\Ba}{\Phi}
\safemath{\Bb}{\Theta}
\safemath{\Opa}{\varphi}
\safemath{\Opb}{\vartheta}
\safemath{\Ta}{T}
\safemath{\Tb}{T}
\safemath{\gaba}{g}
\safemath{\gabb}{h}

\newcommand{\Id}[1]{\mathbf{I}_{#1}}

\safemath{\hcoh}{\mu_{\mathscr{H}}}
\safemath{\coh}{\mu}  
 
\safemath{\cum}{\nu}  
\safemath{\bcoh}{\mu_{\mathrm{B}}}  
\safemath{\bsub}{\nu} 
\safemath{\bcum}{\nu_{\mathrm{B}}}  
\safemath{\borth}{\xi_{\mathrm{B}}} 

\safemath{\hmu}{\hat{\mu}}

\safemath{\smin}{\sigma_{\text{min}}}
\safemath{\smax}{\sigma_{\text{max}}}
\safemath{\wmin}{\omega_{\text{min}}}
\safemath{\wmax}{\omega_{\text{max}}}
\safemath{\lmin}{\lambda_{\text{min}}}
\safemath{\lmax}{\lambda_{\text{max}}}

\newcommand{\sldots}{...}

\safemath{\range}{\mathcal{R}}

\safemath{\es}{\eps_{\mathcal{S}}}

\safemath{\eu}{\eps_{\mathcal{U}}}
\safemath{\ev}{\eps_{\mathcal{V}}}
\safemath{\FT}{\mathcal{F}}
\safemath{\PS}{\proj_{\mathcal{S}}}
\safemath{\PSc}{\pi_{\mathcal{S}^c}}
\safemath{\SI}{\mathfrak{S}}

%% file: intro.tex
\section{Introduction}
The sparse signal recovery literature is vast and has evolved along several threads with  recent focus mostly on probabilistic results.
This  paper constitutes an attempt to consolidate  the numerous coherence-based recovery results available in the literature.
More specifically, we  formulate a single theory that applies to finite- and infinite-dimensional Hilbert spaces, in combination with sparsity defined as the (finite) number of (possibly infinite-dimensional) subspaces participating in a signal's representation.
The general coherence-based recovery thresholds we find contain the known thresholds in the following settings as special cases:
\begin{inparaenum}[(i)]
\item sparse signals in finite-dimensional spaces  \cite{Donoho1989,Donoho2001,Elad2002,Donoho2003}, 
\item block-sparse signals  \cite{Eldar2010,Boufounos2011},
\item multi-band signals  \cite{Feng1996,Bresler2008,Mishali2009a},
\item signals  in shift-invariant  spaces \cite{Eldar2009g}, and
\item signals in finite unions of finite or infinite-dimensional subspaces \cite{Lu2008,Blumensath2009a,Eldar2009d}.
\end{inparaenum}
In addition, we improve upon the thresholds in \cite{Eldar2010} and we generalize the uncertainty relation in \cite{Eldar2009g}.
We introduce suitable generalizations of \mbox{\PZ-minimization} \cite{Donoho2001}, basis pursuit \cite{Donoho2001}, and orthogonal matching pursuit \cite{Tropp2004}.
Finally, we indicate how the results on signal separation reported in \cite{Kuppinger2012,Studer2012} can be extended to the general  Hilbert space setting considered here. 

Key to our results are definitions of coherence \cite{Donoho2001} and mutual coherence \cite{Elad2002,Studer2012} that work for our general setting.
Based on these definitions, we obtain a general kernel uncertainty relation which is then used to establish  general recovery thresholds.
Similarly our definition of mutual coherence paves the way to a general uncertainty relation that yields fundamental limits on how sparse a signal in a general Hilbert space can be under two different representations.
All theorems in this paper are given without proof. 

\subsubsection*{Notation}
\label{sec:notation}
Lowercase boldface letters stand for column vectors and uppercase boldface letters designate matrices.  
For a vector $\va$, the $k$th element is written $a_k$.  
For the matrix $\bA$,  $\bA^\adj$ is its conjugate transpose, its $k$th column is written $\va_k$, and the entry in the $k$th row and $\ell$th column is denoted by~$A_{k,\ell}$.
The spectral norm of $\bA$ is $\norm{l2l2}{\bA}$, $\smin(\matA)$ and $\smax(\matA)$ are the minimum and maximum singular value of $\matA$, respectively. 

\Ha and \Hb are  Hilbert spaces equipped with the norm $\norm{Ha}{\cdot}$ and $\norm{Hb}{\cdot}$, respectively, and \Ha has direct sum decomposition \cite[Ch.~5.20]{Naylor2000} $\Ha = \bigoplus_{i=1}^\subn \sub{\Ha}{i}$ where $\subn<\infty$.
We define $\sub{\sig}{i}$ to be the canonical projection of $\sig$ onto $\sub{\Ha}{i}$.
For $v\in\Ha$, $\norm{Ha0}{\sig} \define \abs{\{ i:\norm{Ha}{\sub{\sig}{i}}>0 \}}$ and $ \norm{Ha1}{\sig} \define \sum_{i=1}^n \norm{Ha}{\sub{\sig}{i}}$. 
We define $\sub{\Ha}{\setS}\define \bigoplus_{s\in\setS} \sub{\Ha}{s}$ and $\sub{\sig}{\setS}$ to be the projection of \sig onto $\sub{\Ha}{\setS}$.
We say that a signal $\sig\in\Ha$ is $\es$-concentrated to the set $\setS$ if $\norm{Ha1}{\sub{\sig}{\setS}} \ge (1-\es) \norm{Ha1}{\sig}$, where $0\le \es \le 1$.  
We define $\ecol_i\in\complexset^N$ to be the all zero vector with a one in the $i$th position.
For an operator $\op\colon \Ha\rightarrow \Hb$ with adjoint $\op^\adj$, $\wmin(\op)\define \inf_{\sig\in\Ha} \norm{Hb}{\op(\sig)}\!/\!\norm{Ha}{\sig}$, $\wmax(\op) \define \sup_{\sig\in\Ha} \norm{Hb}{\op(\sig)}\! / \! \norm{Ha}{\sig}$, and $\ker(\op) \define \{\sig\in\Ha:\op(\sig)=0\}$.
For $\alpha\in\reals$, we set $\pos{\alpha} \define \max\{0,\alpha\}$.
The cardinality of a set $\setS$ is denoted as $\abs{\setS}$.
The  Fourier transform operator is written $\FT$.


%% file: sampling.tex
\section{Signal and Sampling Model}
\label{sec:sig_model}
Let $\Ha$ and $\Hb$ be Hilbert spaces, with dimensions $\dima$ and $\dimb$, respectively,  possibly infinite.  
Assume that $\Ha=\bigoplus_{i=1}^n \sub{\Ha}{i}$, $\subn<\infty$, and  set $d_i=\dim(\sub{\Ha}{i})$. 
We describe the sampling of  signals in \Ha through the application of a bounded linear operator $\samp\colon\Ha\rightarrow\Hb$, which we call a \emph{sampling}  operator.
With $\samp$ we associate the operators $\sampcol_i\colon\sub{\Ha}{i}\rightarrow\Hb$, for $i=1,\sldots,\subn$, obtained by restricting the action of \samp to the subspace $\sub{\Ha}{i}$. 
 It follows from the linearity of $\samp$ that $ \samp(\sig) = \sum_{i=1}^\subn \sampcol_i(\sub{\sig}{i})$.
We require that each $\sampcol_i$ be injective. 

For  $N=\sum_{i=1}^n d_i < \infty$,  the action of \samp can be represented through a matrix  $\dict\in\complexset^{\dimb \times \dima}$ according to $\samp(\sigv) = \dict\sigv$, $\sigv\in\complexset^\dima$.
Taking $\b{\dict}{i}= [\,\dcol_{i_1} \cdots\, \dcol_{i_{d_i}}\,]$ to be the set of columns of \dict that correspond to  $\sub{\Ha}{i}$ we have $\sampcol_i(\sub{\sigv}{i}) = \b{\dict}{i}\sub{\sigv}{i}$, for $i=1,\sldots,n$.
%


%% file: defns_ur.tex
\section{Definitions of Coherence}
\label{sec:coh_ur}
Key to our results are definitions of coherence, mutual coherence, and spark for  general sampling operators. 

\begin{defn}[Hilbert space coherence]
Let $\Ha$ and $\Hb$ be  Hilbert spaces and let $\samp\colon\Ha\rightarrow\Hb$ be a sampling operator. 
We define the \emph{Hilbert space coherence of \samp}as\footnote{By assumption the operators $\sampcol_i$ are injective, hence $\wmin(\sampcol_i)>0$.} 
\begin{align}
   \hcoh = \hcoh(\samp) \define \max_{\substack{i,j,  i\neq j}} \frac{\wmax\lefto( \sampcol_i^\adj \sampcol_j \right)}{\wmin^2\lefto( \sampcol_i\right)}. \label{eq:hcoh}
\end{align}
\end{defn}
We  can  interpret $\hcoh(\samp)$ as a measure of closeness of the subspaces $\sub{\Ha}{i}$ under the action of \samp.

\begin{defn}[Mutual Hilbert space coherence]
Let $\Ha_1$, $\Ha_2$, and $\Hb$ be Hilbert spaces and let $\samp\colon\Ha_1\rightarrow\Hb$ and $\sampb\colon\Ha_2\rightarrow \Hb$ be sampling operators.  
We define the \emph{mutual Hilbert space coherence of \samp and \sampb}as
\begin{align}
   \hcoh(\samp,\sampb) \define \max_{\substack{i,j}} \frac{\wmax\lefto( \sampcol_i^\adj \sampbcol_j \right)}{\wmin\lefto( \sampcol_i\right) \wmin\lefto( \sampbcol_j\right)}. \label{eq:mutualcoh}
\end{align}
\end{defn}
The mutual Hilbert space coherence extends the definition of mutual coherence in \cite{Elad2002,Studer2012}. 
The setting of \cite{Elad2002,Studer2012} is recovered as follows.
Let $\Ha_1=\complexset^{\dima_1}$, $\Ha_2=\complexset^{\dima_2}$, and $\Hb=\complexset^\dimm$. 
Represent the sampling operators $\samp\colon\Ha_1\rightarrow\Hb$ and $\sampb\colon\Ha_2\rightarrow \Hb$ by the matrices $\matA$ and $\matB$, respectively, so that $\samp(\sigv)=\matA\sigv$ and $\sampb(\vec{u})=\matB\vec{u}$.
Then, we have 
\begin{align*}
  \hcoh(\samp,\sampb) &= \max_{\substack{i,j}} \frac{\norm{l2}{\acol_i^\adj \bcol_j}}{\norm{l2}{\acol_i}\norm{l2}{\bcol_j}} \stackrel{(a)}{=} \max_{\substack{i,j}} \abs{\scalprod{}{\acol_i}{\bcol_j}} = \coh_m,
\end{align*}
where $\coh_m$ is the mutual coherence as specified in \cite{Studer2012}, and (a) follows since in \cite{Studer2012} $\matA$ and $\matB$ are assumed  to have columns with unit $\ell_2$-norm.

We will also need a general definition of spark \cite{Donoho2003,Gribonval2003}.
\begin{defn}[Hilbert space spark]
Let $\Ha$ and $\Hb$ be  Hilbert spaces and let $\samp\colon\Ha\rightarrow\Hb$ be a sampling operator.  Then
\begin{align}
    \spark(\samp) \define \min_{\substack{\sig \in \ker(\samp) \setminus\{ {0} \} }} \norm{Ha0}{\sig}. \label{eq:spark}
\end{align}
\end{defn}
The spark of a sampling operator is the smallest number of subspaces that a non-zero signal $\sig\in\Ha$ in $\ker(\samp)$ can occupy.


%% file: recovery.tex
\section{Recovery Thresholds} 
With our general  definitions of coherence and spark, the general recovery thresholds below follow without difficulties. 
We start with a general kernel uncertainty relation.
\begin{thm}[Kernel uncertainty relation] 
\label{thm:up}
Let $\samp\colon\Ha\rightarrow\Hb$ be a sampling operator with Hilbert space coherence $\hcoh(\samp)$.  
Let $\sig\in\Ha$ be $\es$-concentrated to $\setS$.  If $\samp(\sig) = 0$, then
\begin{align}
   \abs{\setS} \ge\left(1-\es\right)\left(1 + \left({\hcoh(\samp)}\right)^{-1}\right).
\end{align}
\end{thm}

We next define two optimization problems for the recovery of a signal $\sig\in\Ha$ from its measurements $\obs=\samp(\sig)\in\Hb$.
The first  one, \HPZ, aims to find the  signal that  explains the given measurements while occupying  the fewest subspaces:
\begin{align}
   (\HPZ) \quad 
       \mini_{\hat{\sig}\in\Ha} \norm{Ha0}{\hat{\sig}} \ 
       \st  \samp(\hat{\sig}) = \obs.
\end{align}
Furthermore, we consider a modified version of basis pursuit:
\begin{align}
   (\HBP) \quad 
       \mini_{\hat{\sig}\in\Ha} \norm{Ha1}{\hat{\sig}} \ 
       \st  \samp(\hat{\sig}) = \obs.
\end{align}

Recovery thresholds for  \HPZ and \HBP can now be derived from the kernel uncertainty relation in \fref{thm:up}.

\begin{thm}
\label{thm:hpz}
If $\sig\in\Ha$ satisfies $\samp(\sig) = \obs$ and 
\begin{align}
  \norm{Ha0}{\sig} &< \spark(\samp)/2, \label{eq:PZ-spark}
\end{align}
then  $\sig$ is the unique minimizer of \HPZ applied to \obs.
\end{thm}

In addition, we have the following bound,  $\spark(\samp) \ge 1+(\hcoh(\samp))^{-1}$, which combined with \fref{thm:hpz} allows us to conclude that \HPZ returns the correct solution  if
\begin{align}
    \norm{Ha0}{\sig} &<  \lefto(1 +\left({\hcoh(\samp)}\right)^{-1} \right)\!/2. \label{eq:recovery}
\end{align}

We next provide  a recovery condition for \HBP.

\begin{thm}
\label{thm:hbp}
If $\sig\in\Ha$ satisfies $\samp(\sig) = \obs$ and \eqref{eq:recovery} holds, 
then \HBP applied to \obs returns the correct solution \sig.
\end{thm}

A commonly used alternative to \BP is  orthogonal matching pursuit (\OMP) \cite{Tropp2004,Davis1997}.  
We next present a Hilbert-space version of \OMP, which we call \Homp.
This  algorithm works by iteratively identifying the subspaces $\sub{\Ha}{i}$ participating in the  representation of \sig and  computes an approximation to  \sig, denoted as $\it{\sig}{i}$, in the $i$th iteration.
The corresponding residual in the $i$th iteration is given by $\it{r}{i} \define \obs - \samp(\it{\sig}{i})$.
The algorithm is initialized with $\it{r}{0} \leftarrow \obs$ and $i\leftarrow 1$, and performs the following steps until $\norm{Hb}{r_i} = 0$:
\begin{enumerate}
\item \label{step:cost} Find  
\begin{align*}
  \ell=\argmax_{\hat\ell} \norm{Ha}{\sampcol_{\hat\ell}^\adj (\it{r}{i-1})}/\wmin(\sampcol_{\hat\ell}).
\end{align*} 
\item Update the list of participating subspaces:~\mbox{$\it{\setS}{i} \leftarrow \it{\setS}{i-1}\cup \{\ell\}$}.
\item Find the best approximation to $\sig$ with support $\it{\setS}{i}$:
\begin{align*}
   \it{\sig}{i} \leftarrow \argmin_{\sigb\in\sub{\Ha}{\it{\setS}{i}} }\norm{Hb}{\obs-\samp_{\it{\setS}{i}}(\sigb)}.
\end{align*}
\item Update the residual and $i$:~$\it{r}{i} \leftarrow \obs - \samp\lefto(\it{\sig}{i}\right)$, $i\leftarrow i+1$.
\end{enumerate}

\begin{thm} \label{thm:homp}
Let $\samp\colon\Ha\rightarrow \Hb$ be a sampling operator. Then \Homp applied to $\obs=\samp(\sig)$ returns the correct solution \sig if \eqref{eq:recovery} is satisfied and will require exactly $\norm{Ha0}{\sig}$ iterations.
\end{thm}

Note that implementing the algorithms mentioned above, when \Ha is infinite-dimensional, is non-trivial. 
Some alternatives to \HBP and \Homp, such as SBR2/4, have been proposed for blind multi-band sampling \cite{Mishali2009a}, which is a special case of our setup.
It is an interesting open problem to extend these algorithms to the  general framework in this paper.


%% file: discussion.tex
\section{Discussion of Recovery Thresholds}
\label{sec:discussion}
We  next show how the recovery thresholds in \cite{Donoho1989,Donoho2001,Elad2002,Donoho2003,Eldar2010,Feng1996,Bresler2008,Mishali2009a,Lu2008} follow from the general recovery threshold \eqref{eq:recovery}.
The results in \cite{Boufounos2011}, which pertain to a generalization of \cite{Eldar2010} allowing for different subspace dimensions,  can be recovered following the same methodology, but this will not be detailed here due to space constraints.

\subsection{Sparse signal recovery}
\label{sec:cs}
The (coherence-based) thresholds in \cite{Donoho1989,Donoho2001,Elad2002,Donoho2003} are recovered as follows.
Set  $\Ha=\complexset^\dima$ and $\Hb=\complexset^\dimm$.
Take the sampling operator \samp to be represented by the matrix $\dict\in\complexset^{\dimm\times\dima}$,  with unit $\ell_2$-norm columns $\dcol_i$.
Take $\sub{\Ha}{i}$ to be the $1$-dimensional subspace spanned by $\ecol_i\in\complexset^\dima$, so that $\dima=\subn$. 
The action of $\sampcol_i\colon\sub{\Ha}{i}\rightarrow \Hb$  is represented by $\sampcol_i(\sub{\sigv}{i}) = \dcol_i \sub{\sigv}{i} = \dcol_i v_i$. 
Since $\wmin(\sampcol_i) = \norm{l2}{\dcol_i}=1$ and $\wmax\lefto( \sampcol_i^\adj \sampcol_j \right) = \abs{\scalprod{}{\dcol_i}{\dcol_j}}$, we get $\hcoh = \max_{\substack{i\neq j}} \abs{\scalprod{}{\dcol_i}{\dcol_j}}$, which is exactly the definition of coherence as introduced in \cite{Donoho2001,Elad2002,Donoho2003}.
The recovery threshold \eqref{eq:recovery} for \HPZ, \HBP, and \Homp (which then reduce to \PZ, \BP, and \OMP, respectively) is thus equal to the corresponding thresholds in \cite{Donoho2001,Elad2002,Donoho2003}.
As an aside the general result \eqref{eq:recovery} shows how dictionaries with unnormalized columns should be treated, specifically what the appropriate measure of coherence is, and what the selection criterion in Step \ref{step:cost} of (\Ha-)\OMP should be.

\subsection{Block-sparsity}
\label{sec:bs}

The results for the block-sparse setting considered in \cite{Eldar2010} are recovered as follows.
Set $\Ha = \complexset^\dima$, $\Hb = \complexset^\dimb$, and $N=nd$, where $d$ is the block size and $n$ is the number of blocks (and hence the number of subspaces $\sub{\Ha}{i}$).
As before,  the sampling operator \samp is represented by the matrix  $\dict\in\complexset^{\dimm\times\dima}$ with unit $\ell_2$-norm columns.
Let $\sub{\Ha}{i}$  be the subspace spanned by $\{\ecol_{(i-1)d+1},\, \sldots,\, \ecol_{id}\}$, and  set $\b{\dict}{i}=[\,\dcol_{(i-1)d+1}\, \cdots\, \dcol_{id}\,]$, so that $\sampcol_i(\sub{\sigv}{i}) = \b{\dict}{i}\sub{\sigv}{i}$.
From \eqref{eq:hcoh} the Hilbert space coherence is 
\begin{align}
  \hcoh(\samp) = \max_{\substack{i,j, j\neq i} } \frac{\smax\lefto(\left(\b{\dict}{i}\right)^\adj \b{\dict}{j} \right)}{\smin^2\lefto(\b{\dict}{i}\right)}. \label{eq:hcoh-block}
\end{align}
We  next show how the  recovery threshold \eqref{eq:recovery}  improves upon that reported in \cite[Thms.~2 and 3]{Eldar2010},  which states that recovery using (\textrm{L-OPT}) \cite[Eq.~32]{Eldar2010} and \BOMP \cite[Sec.~IV-A]{Eldar2010} (our \HBP and \Homp, respectively), is successful if $\norm{Ha0}{\sigv} <  \lefto(1+{\hmu}^{-1}\right)\!/2$.
Here
\begin{align*}
  \hmu &\define \frac{d\bcoh(\dict)}{1-(d-1)\bsub}\\
     \bcoh = \bcoh(\dict) & \define \max_{\substack{i,j, j\neq i} } \frac { \smax\lefto( \lefto(\b{\dict}{i}\right)^\adj \b{\dict}{j}\right)}{d}  \\
     \bsub =\bsub(\dict) &\define \max_{\ell=1,\ldots,\subn} \ \  \max_{\substack{i,j, j\neq i}} \abs{\left(\b{\dict}{\ell}_i\right)^\adj \b{\dict}{\ell}_j},\\ 
\end{align*}
and $\b{\dict}{\ell}_i$ is the $i$th column  of $\b{\dict}{\ell}$.
The following steps establish that $\hcoh\le \hat{\coh}$, thereby proving our claim\footnote{It is possible that $\hat\mu<0$ and since $\hcoh(\samp)$ is always non-negative, we do not have $\hcoh(\samp)\le \hat\mu$ in this case.  However, in this instance \cite[Thms.~2 and 3]{Eldar2010} say that we cannot guarantee the recovery of any signal, but the right-hand side of \eqref{eq:recovery} is positive, thus trivially improving upon the recovery thresholds in \cite[Thms.~2 and 3]{Eldar2010}.}
\begin{align*}
    \hcoh(\samp)     &\le \frac{\max_{j\neq i}\smax\lefto({\left(\b{\dict}{i}\right)^\adj \b{\dict}{j}}\right)} {\min_k \smin\lefto(\left(\b{\dict}{k}\right)^\adj \b{\dict}{k}\right)} \\
    &\stackrel{(a)}{\le} \max_{i,j,j\neq i} \frac{\smax\lefto({\left(\b{\dict}{i}\right)^\adj \b{\dict}{j}}\right)} {\pos{1-(d-1)\bsub}} = \hat{\mu}, 
\end{align*}
where we applied the \Gersgorin disc theorem \cite[Th.~6.1.1]{Horn1990} in~(a).
When $\left(\b{\dict}{i}\right)^\adj\!\b{\dict}{i} = \Id{d}$, for all $ i$, we have $\hcoh=\hat{\coh}$, but one can easily find examples where the strict inequality $\hcoh < \hat{\coh}$ holds.

\subsection{Multi-band signals}
We  next show how our results apply to  sparse multi-band signals as considered in \cite{Bresler2008,Mishali2009a,Venkataramani1998,Mishali2010a}. 
Let \Ha be the space of functions band-limited to the interval $[0, 1/T)$ and for a signal $\sig\in\Ha$, let $V$ be its Fourier transform.
For simplicity of exposition, assume that the interval $[0,1/T)$, is divided into $\subn$ disjoint intervals $\setI_1,\sldots,\setI_\subn$, with $\setI_i=[(i-1)/(nT), i/(nT))$, $i=1,\sldots,\subn$.
Define the subspaces $\sub{\Ha}{i} = \set{\sig\in\Lt(\reals)\colon V(f) = 0,\ \text{for all}\ f \notin \setI_i}$.
Thus, for a signal $\sig\in\Ha$, the sparsity level $\norm{Ha0}{\sig}$ is the number of frequency bands $\setI_i$ occupied by $V$.

We next demonstrate how the multi-coset sampling scheme of \cite{Feng1996,Bresler2008} can be analyzed in our framework.
Multi-coset sampling maps the signal $\sig$ to  $\subm\le \subn$  sequences $\sub{\obs}{k}$ as follows:
\begin{align*}
   \sub{\obs}{k}_\ell = \sig(\ell \subn T + k T), \quad k=1,\ldots,\subm,\ \ell\in\integers.
\end{align*}
To obtain an explicit characterization of the corresponding sampling operator \samp we  will work in the frequency domain.
The  Fourier transform of $\sub{\obs}{k}$  is given by
\begin{align*}
  \sub{Z}{k}\lefto(f\right) &= \frac{1}{\subn T}\sum_{\ell=1}^\subn V\lefto(f+\frac{\ell}{\subn T}\right)  e^{2\pi i k \ell /\subn} \\
&= \frac{1}{\subn T}  \sum_{\ell=1}^\subn \sub{V}{\ell}(f) \, e^{2\pi i k \ell /\subn}  = \sum_{\ell=1}^\subn \lambda_{k,\ell}  \sub{V}{\ell} (f),
\end{align*}
where $\lambda_{k,\ell} =  (\subn T)^{-1} \exp\lefto( 2\pi ik {\ell}/{n}  \right)$ and $\sub{V}{\ell}(f)=V(f+\ell/(nT))$.
Then, the action of the sampling operator, $\samp\colon\Ha\rightarrow \Hb$, can  be represented in terms of the continuously parametrized linear system of equations
\begin{align}
  &\begin{pmatrix}
    \sub{Z}{1}(f)   \pmatspace \\
 \vdots \ppmatspace\\
    \sub{Z}{\subm}(f) \\
  \end{pmatrix} 
  = 
    \underbrace{\begin{pmatrix}
        {\lambda}_{1,1}  & {\lambda}_{1,2}  & \cdots & {\lambda}_{1,\subn} \\
        \vdots & \vdots & \ddots & \vdots\\
        {\lambda}_{\subm,1} & {\lambda}_{\subm,2}  & \cdots & {\lambda}_{\subm,\subn} 
    \end{pmatrix} }_{\define\mat\Lambda}
\underbrace{\begin{pmatrix}
\sub{V}{1}(f)    \pmatspace\\
\vdots \ppmatspace\\
\sub{V}{\subn}(f)
\end{pmatrix}}_{\define\widetilde{V}(f)}, \label{eq:mb-coset}
\end{align}
for $f\in[0,1/(nT))$. 
We have thus established a finite-dimensional continuously indexed matrix representation of \samp\cite{Naylor2000}. 
Based on this insight, we next show that
\begin{align}
  \spark(\samp) &= \spark(\mat\Lambda)=\subm,  \label{eq:mc-spark}
\end{align}
and
\begin{align}
\hcoh(\samp) &= \hcoh(\mat\Lambda), \label{eq:mc-coh}
\end{align}
which means that we can  reduce the computation of  Hilbert space spark and Hilbert space coherence of an infinite-dimensional operator to that of a finite matrix that does not depend on $f$.
Since \eqref{eq:mb-coset} holds for all $f\in[0,1/(nT))$, for $\sig$ to lie in the kernel of \samp, $\widetilde{V}(f)$ must be in $\ker(\mat\Lambda)$ for each $f\in[0,1/(nT))$.  One can then show that this implies that $\spark(\samp)=\spark(\mat\Lambda)$.
The second equality in \eqref{eq:mc-spark} follows since $\mat\Lambda$ consists of the first $\subm$ rows of the $n\times n$ DFT matrix and hence $\spark(\mat\Lambda) = \subm$ \cite{Venkataramani1998}.

To prove \eqref{eq:mc-coh},  note that for $u\in\sub{\Ha}{i}$ with Fourier transform $U$, $\sampcol_i\colon\sub{\Ha}{i}\rightarrow \Hb$ is given by the matrix representation
\begin{align*}
  \sampcol_i\lefto(U\right)(f) \define U(f)\begin{pmatrix} \lambda_{1,i}  \\ \vdots \\ \lambda_{m,i}  \end{pmatrix},
\end{align*}
and has adjoint 
\begin{align*}
  \sampcol_i^\adj\lefto(X\right)(f) =\sampcol_i^\adj\lefto(\begin{pmatrix} \sub{X}{1}\pmatspace\\ \vdots \ppmatspace \\ \sub{X}{m} \end{pmatrix} \right)(f)  \define \left(\sum_{k=1}^m \lambda_{k,i}^*\sub{X}{k}\right)\!(f),
\end{align*}
where
\begin{align*}
  X = \begin{pmatrix} \sub{X}{1} \pmatspace\\  \vdots\ppmatspace \\ \sub{X}{m} \end{pmatrix}\in \Hb.
\end{align*}

Hence, for $u\in\sub{\Ha}{\ell}$ with Fourier transform $U$, we have
\begin{align}
    \norm{Ha}{\sampcol_j^\adj \sampcol_\ell(U)} &= \norm{Ha}{\sum_{i=1}^m \lambda_{i,j}^* \lambda_{i,\ell} U}   = \abs{\sum_{i=1}^m \lambda_{i,j}^* \lambda_{i,\ell} } \norm{Ha}{U} \nonumber \\
&=  \abs{\mat{\lambda}_j^\adj \mat{\lambda}_\ell } \norm{Ha}{U}, \label{eq:ppnorm}
\end{align}
where $\mat{\lambda}_j$ is the $j$th column of $\mat\Lambda$.
Since \eqref{eq:ppnorm} holds for all $U$, it follows that 
\begin{align*}
  \wmin(\sampcol_j^\adj \sampcol_\ell) = \wmax(\sampcol_j^\adj \sampcol_\ell) = \abs{\mat{\lambda}_j^\adj \mat{\lambda}_\ell },
\end{align*}
and hence
\begin{align*}
    \hcoh(\Phi) &= \max_{\substack{j, \ell, j\neq \ell}  } \frac{\wmax({ \sampcol_j^\adj \sampcol_\ell})}{\wmin(\sampcol_\ell^\adj \sampcol_\ell) }=  \max_{\substack{j, \ell, j\neq \ell}  } \frac{\abs{\mat{\lambda}_j^\adj \mat{\lambda}_\ell}}{\norm{l2}{\mat{\lambda}_\ell}^2}
= \hcoh(\mat{\Lambda}).
\end{align*}

From \cite{Feng1996,Mishali2009a} we know that to recover a multi-band signal with bandwidth $s/(nT)$ (and with unknown spectral occupancy), it is necessary to sample at a rate $f_s=m/(nT) \ge 2s/(nT)$.
\fref{thm:hpz} implies that uniqueness of \HPZ recovery is guaranteed for multi-band coset sampling if  $\spark(\samp)/2=m/2>s$.  
Hence, sampling at rate at least $2s/(nT)$ is also sufficient to recover an $s$-sparse signal and recovery of the (multi-coset sampled)  signal can be achieved through  \HPZ.

\subsection{Relation to further results} 
\fref{thm:hpz} in this paper implies \cite[Prop.~4]{Lu2008} and \cite[Eq.~(23)]{Lu2008} with the observation that the generalized Gram matrix in \cite[Eq.~(17)]{Lu2008} plays the role of the sampling operator~$\samp$ in our framework.
Our \fref{thm:hpz} also implies \cite[Th.~2.2]{Blumensath2009a}.


%% file: uncertainty.tex
\section{Uncertainty Relations and Signal Separation}
\label{sec:gen_up}

Another  thrust in the sparse signal recovery literature deals with the recovery of sparsely corrupted signals \cite{Studer2012}.
The main tool underlying this line of work is an uncertainty relation that sets a limit on how sparsely a given signal can be represented concurrently in two different dictionaries \cite{Donoho1989,Kuppinger2012,Studer2012}.
We  next formulate a Hilbert space version of this uncertainty relation, which is then used to recover and generalize  results in  \cite{Eldar2009g} and \cite{Studer2012}.

\begin{thm}[Uncertainty relation] \label{thm:gup}
Let~$\Ha_1,\Ha_2$, and $\Hb$ be Hilbert spaces .
and let $\samp\colon\Ha_1\rightarrow \Hb$ and $\sampb\colon\Ha_2\rightarrow \Hb$ be  sampling operators.
Let $\sigb\in\Ha_1$ and $\siga\in\Ha_2$ be signals that are $\eu$- and $\ev$-concentrated to the sets $\sigbsup$ and $\sigsup$, respectively, and assume that $\samp(\sigb) = \sampb(\siga)$.
Then, we have
\begin{align}
\abs{\sigbsup} \!\abs{\sigsup} &\ge  \frac{1}{\hcoh^2(\samp,\Psi)}\lefto[\lefto(1-\eu\right)\lefto(1+\hcoh(\samp)\right)-\abs{\setU}\hcoh(\samp)\right]^+ \nonumber \\
&\ \ \times\lefto[\lefto(1-\ev\right)\lefto(1+\hcoh(\sampb)\right)-\abs{\setV}\hcoh(\sampb)\right]^+. \label{eq:gup}
\end{align}
\end{thm}

Remark:~\cite[Th.~1]{Studer2012} can be recovered from \fref{thm:gup} by noting that \samp and \sampb play the role of the  dictionaries $\matA$ and $\matB$, respectively, as used in \cite{Studer2012}. 
Then $\samp(\vecu)=\sampb(\vecv)$ becomes $\matA\vecu = \matB\vecv$ and \cite[Th.~1]{Studer2012} follows since $\hcoh(\samp,\sampb) = \mu_m$, $\hcoh(\samp) =\mu_a$, and $\hcoh(\sampb)=\mu_b$, with $\mu_m,\mu_a$, and $\mu_b$ as defined in \cite{Studer2012}.

\subsection{Shift-invariant spaces}
\label{sec:sispace}
We next show how \fref{thm:gup} can be used to recover  \cite[Th.~1]{Eldar2009g}.
Consider the shift-invariant space 
\begin{align}
    \SI_{\sia} \define \set{ \obs \colon \obs(t) = \!\!\! \sum_{\substack{i=1,\ldots,n_1\\k\in\integers}} \!\!\! \sub{\sig}{i}_k \sia_i(t-k\Ta), \sub{\sig}{i}\in\lt,\ \forall i}, \label{eq:sia}
\end{align}
with $n_1$ generators $\sia_i\in\Lt(\reals)$ and $\norm{l2}{\sia_i}=1$, for all $i$.
Set $\Ha_1$ to be the space of vector sequences 
\begin{align}
  v = \begin{pmatrix} \sub{v}{1} \pmatspace\\  \vdots\ppmatspace \\ \sub{v}{n_1} \end{pmatrix},   \label{eq:l2vs}
\end{align}
with $\sub{v}{i}\in\lt$, for all $i$.
Define the operator $\sampcol_i\colon\lt\rightarrow \SI_\sia$ by
\begin{align}
  \sampcol_i\lefto(\sub{\sig}{i}\right) \define \sum_{k\in\integers} \sub{\sig_k}{i} \sia_i(\cdot - k \Ta), \label{eq:sisampcol}
\end{align}
with adjoint  $\sampcol_i^\adj\colon\SI_\sia \rightarrow \lt$ given by $ \sampcol_i^\adj(z) = \left\{ \scalprod{}{z(\cdot) }{\sia_i(\cdot - \ell \Ta)}\right\}_{\ell\in\integers}$.
The sampling operator\footnote{In this case, the sampling operator rather behaves like an interpolation operator as it maps a sequence to a continuous-time signal, but to maintain consistency with the rest of the paper we still refer to it as a sampling operator.} $\samp\colon\Ha_1\rightarrow \SI_\sia$ is then given by
\begin{align}
  \samp(\sig) &\define \sum_{i=1}^{\subn_1} \sum_{k\in\integers} \sub{\sig_k}{i} \sia_i(\cdot - k \Ta) = \sum_{i=1}^{\subn_1} \sampcol_i\lefto(\sub{\sig}{i}\right). \label{eq:sisamp}
\end{align}
A signal $\sig\in\Ha$ is $s$-sparse  if at most $s$ of the sequences $\sub{\sig}{i}$ in \eqref{eq:sisamp} are non-zero, i.e., if $\norm{Ha0}{\sig}\le s$, and in the terminology of \cite{Eldar2009g}, $\norm{Ha0}{\sig}$ is the number of active generators.

Now let us consider a set of $n_2$ generators  $\sib_i\in\Lt(\reals)$ where $\norm{l2}{\sib_i}=1$, for all $i$, and the space 
\begin{align*}
\SI_\sib \define \set{ \obs \colon \obs(t) = \!\!\! \sum_{\substack{i=1,\ldots,n_2\\k\in\integers}} \!\!\! \sub{\sig}{i}_k \sib_i(t-k\Ta), \sub{\sig}{i}\in\lt,\ \forall i}.
\end{align*}
Let $\Ha_2$  be the space of  vector sequences, as in \eqref{eq:l2vs}, but with $n_1$ replaced by $n_2$, and define the operators $\Opb_i\colon \lt \rightarrow \SI_\sib$ and $\Bb\colon\Ha_2\rightarrow \SI_\sib$  as in \eqref{eq:sisampcol} and \eqref{eq:sisamp}, respectively, with $\sia_i$ replaced by $\sib_i$. 
Suppose that $\obs=\Ba(v) = \Bb(u)$. 
We now establish a limit on the sparsity of $u$ and $v$.

Following \cite{Eldar2009g} the generators will be assumed to satisfy:
\begin{align*}
    \scalprod{}{\sia_i(\cdot-kT)}{\sia_j(\cdot-\ell T)} &= \scalprod{}{\sib_i(\cdot-kT)}{\sib_j(\cdot-\ell T)}\\
      &=\begin{cases} 1 & \text{if } i=j \text{ and } k=\ell \\ 0 & \text{otherwise.}
    \end{cases}
\end{align*}
Then $\norm{l2}{\sampcol_i(\sub{\sig}{i})}=\norm{l2}{\sub{\sig}{i}}$, for all $i$, and for all $\sub{\sig}{i}\in\lt$, hence $\wmin(\sampcol_i) = 1$, for all $i$, and similarly $\wmin(\Opb_\ell) = 1$, for all $\ell$.
For $i\neq j$ and $\sub{\sig}{j}\in\lt$, we have
\begin{align*}
&  \norm{l2}{\Opa_i^\adj \Opa_j\lefto(\sub{\sig}{j}\right)} = \sum_{k\in\integers} \abs{\scalprod{}{\Opa_j\lefto(\sub{\sig}{j}\right)}{\sia_i(\cdot - k T)}}^2 \\
&\qquad= \sum_{k\in\integers} \abs{\scalprod{}{ \sum_{\ell\in\integers} \sub{\sig}{j}_\ell \sia_j(\cdot-\ell T)}{\sia_i(\cdot - k T)}}^2 \\
&\qquad=\sum_{\substack{k\in\integers\\ \ell\in\integers}} \abs{\sub{\sig}{j}_\ell}^2 \underbrace{\abs{\scalprod{}{  \sia_j(\cdot-\ell T)}{\sia_i(\cdot - k T)}}^2}_{=0} = 0,
\end{align*}
and similarly $\norm{l2}{\Opb_i^\adj \Opb_j\lefto(\sub{\sigb}{j}\right)} =0$, for all $\sub{\sigb}{j}\in\lt$.
Therefore, $\hcoh(\Ba)=\max_{i\neq j} \wmax(\Opa_i^\adj \Opa_j) = 0 $ and similarly $\hcoh(\Bb)=0$.
This gives
\begin{align*}
  \hcoh(\Ba,\Bb) &= \max_{i, j} \frac{\wmax\lefto(\Opa_i^\adj \Opb_j\right)}{\wmin\lefto(\Opa_i\right)\wmin\lefto(\Opb_i\right)} = \max_{i, j}  \wmax(\Opa_i^\adj \Opb_j).
\end{align*}
The uncertainty relation \eqref{eq:gup} hence reduces to
\begin{align}
     \norm{Ha0}{\sigb}\norm{Ha0}{\sig}\, {\hcoh^2(\Ba,\Bb)} \ge 1, \label{eq:si_orth_res}
\end{align}
where we assume perfect concentration (since this is the case considered in \cite{Eldar2009g}), i.e., $\eu=\ev=0$.
We  now show that \eqref{eq:si_orth_res} is the uncertainty relation in \cite[Th.~1]{Eldar2009g}, which, in our notation, is given by \eqref{eq:si_orth_res} but with $\hcoh(\Ba,\Bb)$ replaced by
\begin{align}
    \coh(\Ba,\Bb) &= \max_{\ell, r}\  \esssup_{\xi\in[0,2\pi)} \abs{R_{\sia_\ell, \sib_r}\lefto(e^{i\xi}\right) }, \nonumber
\end{align}
where
\begin{align}
    R_{\sia_\ell, \sib_r}\lefto(e^{i\xi}\right) &\define \FT\lefto\{ \big\{\!\!\scalprod{}{\sib_r(\cdot)}{\sia_\ell(\cdot- kT)}\!\!\big\}_{k\in\integers} \right\}\!\lefto(e^{i\xi}\right). \nonumber
\end{align}
It therefore suffices to prove that $\wmax\lefto(   \Opa_\ell^\adj \Opb_r\right)=\esssup_{\xi\in[0,2\pi)} \abs{R_{\sia_\ell, \sib_r}\lefto(e^{i\xi}\right) } $.
For $\sub{\sig}{r} \in \lt$, we have
 \begin{align*}
     \Opa_\ell^\adj \Opb_r\lefto(\sub{\sig}{r}\right)      &= \left\{ \sum_{k\in\integers} \sub{\sig}{r}_k\scalprod{}{ \sib_r(\cdot -kT)}{\sia_\ell(\cdot -m T)} \right\}_{m\in\integers}\\
     &= \left\{ \sum_{k\in\integers} \sub{\sig}{r}_k \Lambda_{m ,k} \right\}_{m\in\integers},
 \end{align*}
where $\Lambda_{m, k} \define \scalprod{}{ \sib_r(\cdot -kT)}{\sia_\ell(\cdot -m T)}$.
Then 
\begin{align*}
    \wmax\lefto(  \Opa_\ell^\adj \Opb_r \right) &= \sup_{\norm{l2}{
\sub{\sig}{r}}=1} \norm{l2}{ \Opa_\ell^\adj \Opb_r(\sub{\sig}{r}) }\\
    &= \sup_{\norm{l2}{\sub{\sig}{r}}=1} \norm{l2}{\mat{\Lambda}\sub{\sig}{r}} = \norm{l2l2}{\mat{\Lambda}}.
\end{align*}
Since $\mat\Lambda$ is a doubly infinite Toeplitz matrix, its operator norm $\norm{l2l2}{\mat{\Lambda}}$ is given by the essential  supremum of the  Fourier transform of a row of $\mat{\Lambda}$ \cite[p.~62]{Grenander1984}.  We therefore have $
  \wmax\lefto(   \Opa_\ell^\adj \Opb_r\right) = \esssup_{\xi\in[0,2\pi)} \abs{R_{\sia_\ell, \sib_r}\lefto(e^{i\xi}\right) }$,
which concludes the proof.

We finally note that our \fref{thm:gup} also applies to nonorthogonal generator sets $\{\sia_i\}$ and $\{\sib_j\}$ with potentially different shift parameters,  thereby extending the uncertainty relation in \cite{Eldar2009g}.
